\documentclass[twocolumn,showpacs,preprintnumbers,amsmath,amssymb]{revtex4}
 
\usepackage{graphicx}
\usepackage{dcolumn}
\usepackage{bm}

\begin{document}
\preprint{??}

\title{Antiproton-Proton Channels in  J$/\psi$ Decays}

\author{ B.Loiseau$^1$}
\author{ S. Wycech$^2$}
\email{loiseau@in2p3.fr}
\email{wycech@fuw.edu.pl}
\affiliation{%
$^1$LPNHE\footnote{Unit\'e de Recherche des Universit\'es Paris 6 et Paris 7, associ\'ee au CNRS}, Groupe Th\'eorie, Universit\'e P. \& M. Curie,
 4 Pl. Jussieu, F-75252 Paris, France\\
 $^2$Soltan Institute for Nuclear Studies,
Warsaw, Poland
}

\date{\today}

\begin{abstract}
An article usually includes an abstract, a concise summary of the work
covered at length in the main body of the article. It is used for
secondary publications and for information retrieval purposes. Valid
PACS numbers may be entered using the \verb+\pacs{#1}+ command.
\end{abstract}


\begin{abstract}
The recent measurements by the BES Collaboration of J/$\psi$ decays into $
\gamma p {\bar p}$ indicate  a strong enhancement at $ p {\bar p}$
threshold  not observed in the decays into  $ \pi^0 p {\bar p}$. 
Is this enhancement due to a $ p {\bar p}$ quasi-bound state or a
baryonium? A natural explanation follows from a traditional
model of $ p \bar{p} $ interactions based on $G$-parity  transformation. 
The observed $ p {\bar p}$ structure is due to a strong 
attraction in the  $^{1}S_{0} $ state, and  
possibly to  a  near-threshold quasi-bound state in 
the  $^{11}S_{0} $ wave.  
\end{abstract}
 

\maketitle
\thispagestyle{empty}


A  point of interest   in the antiproton interactions
is the question of  existence or non-existence  
of exotics  in the  nucleon-antinucleon ($N \bar{N}$) systems: 
quasi-bound, virtual,  resonant, multiquark or baryonium states~\cite{kle02}.
Such states, if located close to the threshold,  may be indicated by 
large  scattering lengths for 
a given spin and isospin state. For this purpose the scattering 
experiments are apparently 
the easiest to perform with a good precision. However,  
a clear separation of   
quantum states  is not  easy.  Complementary measurements of    the 
X ray transitions in antiprotonic hydrogen are  
useful to select some partial waves.  
These are  particularly valuable when the  fine structure of levels 
is resolved. 
Such a resolution  has been  achieved for the 1S states~\cite{aug99}   
and partly for  the 2P states~\cite{augnp99}.  
Another method   to reach selected states are  formation experiments. 
In this way a resonant-like behavior was recently   observed  
by BES Collaboration in the radiative  decay 
\begin{equation}
\label{radiative} 
 J/\psi \rightarrow \gamma p\bar{p}
\end{equation}
close to the $ p\bar{p} $ threshold~\cite{bai03}. 
On the other hand, a clear threshold suppression is seen in the  
pionic  decay channel 
\begin{equation}
\label{pionic} 
 J/\psi \rightarrow  \pi^0  p\bar{p}. 
\end{equation}
To understand  better the nature of the enhancement,  one should 
look into the $ p\bar{p} $  sub-threshold-energy region. 
This  may be achieved, indirectly,  in the  $ \bar{p}d $ 
low-energy scattering or $ \bar{p}d $ atoms. Such  atomic 
experiments were performed,  but  the fine structure resolution 
has not been achieved so far~\cite{got99}. 

The purpose of this letter   is to discuss  the  physics of 
slow $p \bar{p}$ pairs produced  in the $J/\psi$ decays. 
The $J^{PC}$ conservation reduces the number of 
$ p {\bar p}$  final states 
 to several  partial waves. These, denoted by $^{2I+1~2S+1}L_{J}$,
  differ by their isospin $I$, spin 
$S$, angular 
momenta $L$ and  total spin $J$. Close to the $p \bar{p}$ threshold,
quite  different behavior of   scattering amplitudes is 
expected in different states.  In the $1S$ state of antiprotonic hydrogen it 
is the   
$ ^{1}S_0 = (^{11}S_{0} +\ ^{31}\!S_{0})/2 $   and 
$ ^{3}S_{1} = (^{13}S_{1} +\ ^{33}\!S_{1})/2 $  waves which are 
studied~\cite{aug99}.
While  atomic experiments  determine  the  scattering lengths, 
the BES experiment allows to extend this knowledge  into a broad 
energy region above the threshold. As will be shown,  
the radiative  $J/\psi$ decay  involves  also  the  
$^{11}S_{0} +\ ^{31}\!S_{0}$ combination.
The understanding of this  and other involved states should be 
based on    
the   experience gained in  studies of   elastic  and 
inelastic $ \bar{N}N$ scattering.  We use the   Paris potential 
model~\cite{par05,par99,par94,par82} for this purpose.

To our present  knowledge,  none of the available related 
works~\cite{sib04,ker04,bug04,gao04,zou04} on the $J/\psi$ 
decays~(\ref{radiative}) and~(\ref{pionic})  have given 
a comprehensive explanation of 
the BES experimental spectra. Only two of these papers~\cite{sib04,bug04} 
compare their results to the data.
The J\"{u}lich $N \bar{N}$ model is used in~\cite{sib04} to show that, 
within the Watson-Migdal 
approach,  the isospin $1$ $S$-wave can reproduce the low energy part 
of the $p \bar{p}$ spectrum in the radiative decay.   The 
same spectrum is fitted with a 
constant scattering length in Ref.~\cite{bug04}. The length obtained 
in this way is larger than the lengths calculated  in potential models.  
In Ref.~\cite{ker04}, more realistic but spin averaged constant 
lengths are shown
to generate some low-energy enhancement in reaction~(\ref{radiative}).  
A $K$-matrix, calculated with the one-pion exchange in 
the Born approximation, is considered in Ref.~\cite{zou04}. 
An enhancement is seen, but  this model is too simple to 
describe the  $N \bar{N}$ interactions. 
The formation mechanisms in the radiative decays are 
discussed qualitatively in  Ref.~\cite{gao04}, where 
the quantum numbers of final states are listed with  
the recommendation to look into decay modes of 
the $p \bar{p}$  systems.

In the present work the following results are obtained. 
The set of allowed final  $p \bar{p}$ states is limited to three  
partial waves in the photon 
channel  and to  two waves in the pion channel. 
Among the three possible   $ p\bar{p} $ states in the 
 $p \bar{p} \gamma$ channel 
one is dominated, at very low energies,  by  the well known  
$p \bar{p}(^{13}P_0)$  resonance, formed as a result of 
attractive  
one-pion exchange forces. However, this state  as well as  
another allowed  $^{3}P_1 $ state cannot explain  the 
experimental   spectrum. 
The final $p \bar{p} \gamma$ state 
is   dominated by  the $p \bar{p}(^1S_0)$ wave. 
A strong attraction 
arises  in this wave as a result of coherent one-pion  and two-pion 
exchange forces. It produces  broad, deeply bound  states,  
difficult to detect. However, the recent version of the 
model~\cite{par05}, adapted to hydrogen atom data,   generates 
a near-threshold state  in  the related  
$p \bar{p}(^{11}S_0)$ wave. This state is about 50~MeV wide 
and  bound by 5~MeV.

In the  $p \bar{p} \pi^0$ decay  channel, isospin being conserved,  two $p \bar{p} $ waves, 
   $^{33}S_1$ and   $^{31}P_1$,  are allowed. These  
indicate distinctly different  threshold behavior. 
The $S$ wave is ruled out by the experiment and the $p \bar{p}(^{31}P_1)$
leads to  a natural explanation of the BES  spectrum. 
These findings  can  be unified in a qualitative model for both  
decay modes.

\textit {The allowed  final  states}.    
The  $J^{PC}$ conservation  limits the  number of  slow $p \bar{p}$ 
final states. 
The latter are understood   as $p \bar{p}$ pairs with  
small $ M_{p\bar{p}}-2m_p$ 
where  $ M_{p\bar{p}}$ is the pair  
invariant mass. The allowed states   are  listed 
in Table~\ref{table1} and a few  possibilities exist 
for each  channel.

\begin{table}
\caption{ States  of low energy  $p \bar{p}$   pairs   allowed in the 
$ J/\psi \rightarrow \gamma p\bar{p}$ and  
$ J/\psi  \rightarrow \pi^0 p\bar{p}$
decays. The first column  gives  the decay  modes to  the specified 
internal state  of the  $ p\bar{p} $ pair. Well established,   
two particle  analogs are  indicated in the second 
column~\protect\cite{PDG02}. The third 
 column gives  $J^{PC}$ for the 
light spectator  particles, photons or pions. The  fourth  column gives 
$J^{PC}$ for the internal $p \bar{p}$ system; 
the last column gives the relative  angular momentum of the light 
particle  vs. the pair.   
$  J^{PC} = 1^{(--)}$ for  $ J/\psi $.  }
\begin{ruledtabular}
\begin{tabular}{lcccc}
Decay mode &   Analog                      &  $J^{PC}[\gamma~\rm{or}~ \pi^0] $    &  $J^{PC}[p \bar{p} ]$   & Relative l      \\
  $\gamma  p \bar{p} (^1S_0)$ & $\gamma  \eta(1444)  $  & $ 1^{--}$    &  $0^{-+}$      &  1              \\
  $\gamma  p \bar{p} (^3P_0)$ & $\gamma  f_0(1710)   $  &  $ 1^{--} $  &  $ 0^{++} $        &  0              \\

\vspace{0.12cm}

  $\gamma  p \bar{p} (^3P_1)$ & $\gamma  f_1(1285)   $  &  $ 1^{--} $  &  $ 1^{++} $        &  0              \\
 $\pi^0   p \bar{p} (^{31}P_1) $ &  &  $ 0^{-+}  $    &  $ 1^{+-} $ &   0              \\
 $\pi^0   p \bar{p} (^{33}S_1) $ & $\pi^0   \rho  $  &  $ 0^{-+}$  & $1^{--}   $     &  1           \\
\end{tabular}
\end{ruledtabular}
\label{table1}
\end{table}

\begin{table*}
\caption{ Radiative couplings  of  low energy  $p \bar{p}$   pairs    
to $\gamma \  J/\psi   $ and  angular distributions  of the 
final photon  with respect 
to the $ J/\psi $ spin direction,  $J$  denotes the  $ J/\psi $ and   
$h$  the $ p\bar{p} $  pair.  The second line   specifies  the
transformation property  of  $ p\bar{p} $ 
pair.  The third line gives  the  invariant couplings and 
the fourth line  presents these in  the $ J/\psi$ center of mass.
$k^{\gamma,J,h}_{\alpha}$ are the four momenta and 
$\epsilon^{\gamma,J,h}_{\alpha}$ the polarization four-vectors of the corresponding 
fields.
The last line  gives the  angular distributions for the three different cases. 
The  photon energy is denoted by $\omega $, that 
 of   $ p\bar{p} $ pair  by   $E_h$ and 
the $ J/\psi $ mass by  $ M_J$. For $E_h= 2m_p: \omega/E_h = 0.46 $. }
\begin{ruledtabular}
\begin{tabular}{cccc}
decay mode             & $\gamma\  p \bar{p} (^1S_0) $        &  $\gamma\  p \bar{p} (^3P_0)$      &   $\gamma\  p \bar{p} (^3P_1)$      \\
 h($p \bar{p}$)  & pseudoscalar & scalar & pseudovector \\
coupling & $ \varepsilon^{\alpha \beta \sigma  \rho} k^{\gamma}_{\alpha} \epsilon^{\gamma}_{\beta}k^{J}_{\sigma}   \epsilon^{J}_{\rho }/M_J $ &  $( k^{J}_{\sigma} \epsilon^{J}_{\rho }- k^{J}_{\rho} \epsilon^{J}_{\sigma}) F^\gamma_{\sigma \rho}/M_J $ &
$ \varepsilon^{\alpha \beta \sigma  \rho} k^{\gamma}_{\alpha} \epsilon^{\gamma}_{\beta}\epsilon^ {J}_{\sigma}   \epsilon^{h}_{\rho } $\\
coupling  in CM & $(\mbox{\boldmath$\epsilon$}^{J}\wedge\mbox {\boldmath$\epsilon$}^{\gamma})\cdot  \mbox{\textbf{k}}^{\gamma} $ &
$  \omega$ \mbox{\boldmath $\epsilon$}$^{J} $. \mbox{\boldmath $\epsilon$}$^{\gamma} $   &
$(\mbox{\boldmath$\epsilon$}^{J}\wedge\mbox {\boldmath$\epsilon$}^{\gamma})\ \cdot$ 
[$\mbox{\textbf{k}}^{\gamma}(\textbf{k}^{h}\cdot \mbox {\boldmath$\epsilon$}^{h})/E^h-\mbox{\boldmath$\epsilon$}^{h}\omega]$ \\
angular  profile & $1+\cos^2\theta$ & $1+\cos^2\theta$ & $ 2\sin^2\theta + [(1+\omega/E^h)/(1-\omega/E^h)] (1+\cos^2\theta)$\\
\end{tabular}
\end{ruledtabular}
\label{table2}
\end{table*}

The BES experiment  provides an   angular distribution for the  
photons.  
The specific situation of  $ e^- e^+$  collision is 
that the   projectiles are polarized perpendicular to the beam 
direction and the $ J/\psi $ spin direction 
follows that of  the beam.  The angular distribution  
can be measured within a limited range. It  offers a hint 
which  in principle    
permits  to further reduce  the number of allowed final  states.  
For completeness, the angular distributions of the  photon   
in the states of  interest is   calculated below. 
The simplest Lorentz invariant  couplings which are also gauge 
invariant are given in Table~\ref{table2}.
The  $ p\bar{p} $ pair is described as a single   
scalar $ ( ^3P_0 )$, 
pseudoscalar $ ( ^1S_0 )$  or pseudovector $ ( ^3P_1 )$ particle. 
This picture  is expected 
to work for the  pairs of small  center of mass (CM) system energies. 
Next,  the appropriate couplings are reduced to the   
$ J/\psi $  CM system and the 
corresponding angular distributions are calculated.
The couplings given in Table~\ref{table2}  follow from Hamiltonians 
that couple the electromagnetic field $f_{\sigma \rho}$ 
to the $J/\psi$ field  
$F_{\sigma \rho}=\partial_\sigma V^J_\rho - \partial_\rho V^J_\sigma$ and 
to the corresponding 
scalar $\varphi^{s}$, pseudoscalar $\varphi^{ps}$ and pseudovector
$\varphi^{pv}_{\sigma}$ fields describing the  low energy  
$p \bar{p}$   pairs.  These  gauge invariant  Hamiltonians are 
$H^{s}= (e \kappa^{s}/2M_J )
\int  f_{\sigma \rho}F^{\sigma \rho}\varphi^{s} $, 
$H^{ps}= (e \kappa^{ps}/ 4M_J)
\int  \tilde{f}_{\sigma \rho}F^{\sigma \rho}\varphi^{ps} $~\cite{gur71} and
$H^{pv}= e \int  f_{\sigma \rho}V^{J}_{\alpha} \varphi^{pv}_{\beta}
\varepsilon^{\sigma  \rho \alpha \beta } $, 
 where $\kappa^{s,ps}$ are the magnetic moments for the transitions. 
The vector nature of the fields, expressed by  
$\nabla^{\beta} \varphi^{pv}_{\beta}=0$,  leads to the 
 results given in Table~\ref{table2}. The first two angular distributions 
are well known~\cite{ren86}. $\theta$  denoting the angle 
between the  $\gamma$  emission and  the beam    direction, 
the angular distributions  $(\cos^2\theta + 1) $  and  
$\sin^2\theta $  were tested against the data in Ref.~\cite{bai03}. 
These have indicated  a  preference for the first choice, 
i.e.  radiative transitions to  $  ^3P_0 $ or $ ^1S_0 $ states.  
However, as can be seen in  Table~\ref{table2},  a transition to  
$  ^3P_1 $ state   is not  excluded.


\textit{Final state interactions}. 
Any multichannel system can  be conveniently parameterized by  a  $K$-matrix 
which guarantees  unitarity of the  description. 
The transition amplitude from a channel $i$ to a two-body channel $f$ may be 
presented in  the form  
\begin{equation}
\label{pilkuhn} 
T_{if}= \frac{A_{if}  }{ 1   +   i  q  A_{ff} }
\end{equation}
where $A_{if}  $ is a transition length,  $A_{ff}  $ is the scattering 
length in the channel 
$f$,  and $ q $ is the momentum in this channel (see e.g.~\cite{pil03}).
Both lengths can  be expressed in terms of energy dependent 
$K$ matrix elements. The  same formalism describes the scattering 
amplitude in the channel $f$ as  
\begin{equation}
\label{channelf} 
T_{ff}= \frac{A_{ff}  }{ 1   +   i  q A_{ff}}. 
\end{equation}
In the process of interest the formation amplitude $A_{if} $ is  unknown, 
 but  $ A_{ff}$ is calculable in $ N {\bar N } $ interaction  models  
constrained  by  other experiments. For slow $ p {\bar p} $ pairs the  final 
state interactions in the 
 $ \pi^0 p {\bar p}$ and   $ \gamma  p {\bar p}$ systems 
are  dominated by  interactions in the  $ p {\bar p} $ sub-system.
A formal manipulation of Eqs.~(\ref{pilkuhn}) and~(\ref{channelf})  yields   
\begin{equation}
\label{watson} 
T_{if}= \frac{A_{if}  }{ A_{ff} } T_{ff} = 
\left ( \frac{A_{if} q^{L} }{ A_{ff}}  \right)   
\left ( \frac{T_{ff} }{q^{L}  } \right )  ,  
\end{equation}
which defines  a quantity 
$C_{if} \equiv  A_{if}  q^{L}/  A_{ff}$. 
For $S$-waves, the standard  final state dominance assumption 
(Watson-Migdal) is equivalent to a  weak energy dependence in $C_{if} $. 
This  is usually true in a  small energy range   
where  the denominator in Eq.~(\ref{pilkuhn}) provides all the energy
dependence. In the  $ p {\bar p}$ states  such an approximation is correct 
for $q$  up to about 0.5~fm$^{-1}$. It fails at  higher momenta 
since   $ A_{ff} $ is  energy dependent.
On the other hand $ A_{if} $ stems  from a  short range $c\bar{c}$ 
annihilation process. The annihilation range  is of  the order 
of $1/m_{c}$~\cite{kro98} and only a  weak energy dependence is  expected 
in $ A_{if} $. 
We assume  $ A_{if} = A_{if}(0)/\left[ 1 + (r_o q)^2 \right ] $ 
with a range parameter $r_o$ well below 1~fm. 
For $P$-wave final states,  the low-energy behavior gives 
$ A_{if}\approx   A^{1}_{if} q  $,   
$ A_{ff}\approx   A^{1}_{ff} q^2 $ where the
 $ A^1_{if} $  are parameters and the $ A^1_{ff} $ the scattering volumes. 
The latter  are  
energy dependent as a result of medium  ranged  $\pi$ exchange forces.
This dependence  is  particularly strong in those waves  which involve 
resonances.  The Watson  approximation is  not appropriate there, 
 and   Eq.~(\ref{pilkuhn}) must be  used. The transition length is 
parameterized as  $ A_{if} = A^{1}_{if}(0) q /\left [1 + (r_o q)^2 \right ]^2 $.  
\begin{figure}
\label{fig1}
\includegraphics*[width=8.5cm]{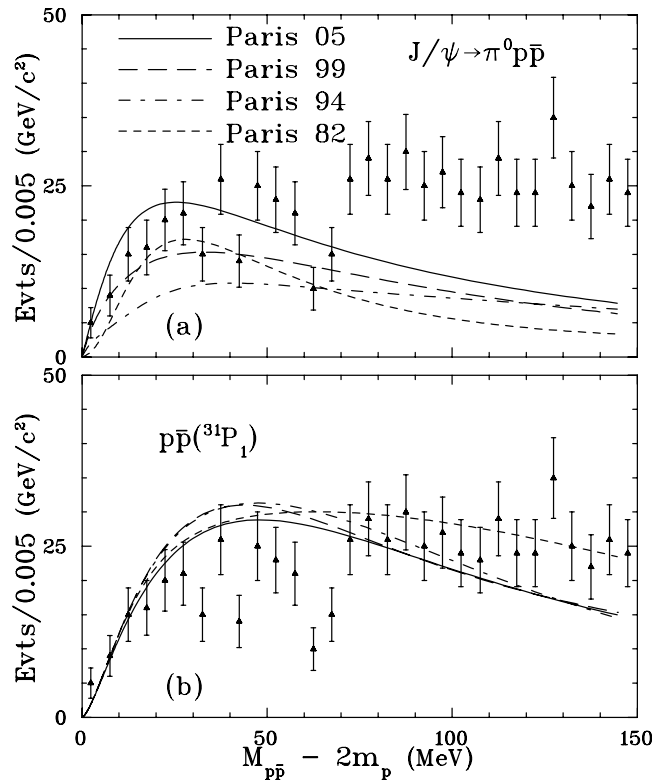}
\caption{ $ \pi^0  p \bar{p}  (^{31}P_1) $ decay channel.
Experimental data have been extracted from Fig.~2(a) 
of Ref.~\protect\cite{bai03}. (a) Final state factor $q\mid T_{ff}/q \mid^2  $
 (Watson approximation).
Constant $C_{if}$ of Eq.~(\ref{watson}) is chosen to fit  the  
low-energy part of the  data. Four versions of the  Paris  potential model~\protect \cite{par05,par99,par94,par82} are used.  
This approximation  fails  for $ M_{p\bar{p}}-2m_p >40$~MeV 
($ q>1$~fm$^{-1}$). 
(b) The  rate $q\mid T_{if}\mid^2  $  of Eq.~(\ref{pilkuhn}). Constant $A^1_{if}(0)$ 
and formation range parameter $r_o=0.55$~fm 
are  chosen to obtain a  good fit to the data. 
All four potentials  give  equivalent fits,
even-though a 118~MeV wide  state bound by  
15~MeV  is  generated   
in  version~\protect \cite{par05} in the $^{31}P_1$ wave.}    
\end{figure}

\begin{figure}
\label{fig2}
 \includegraphics*[width=8.5cm]{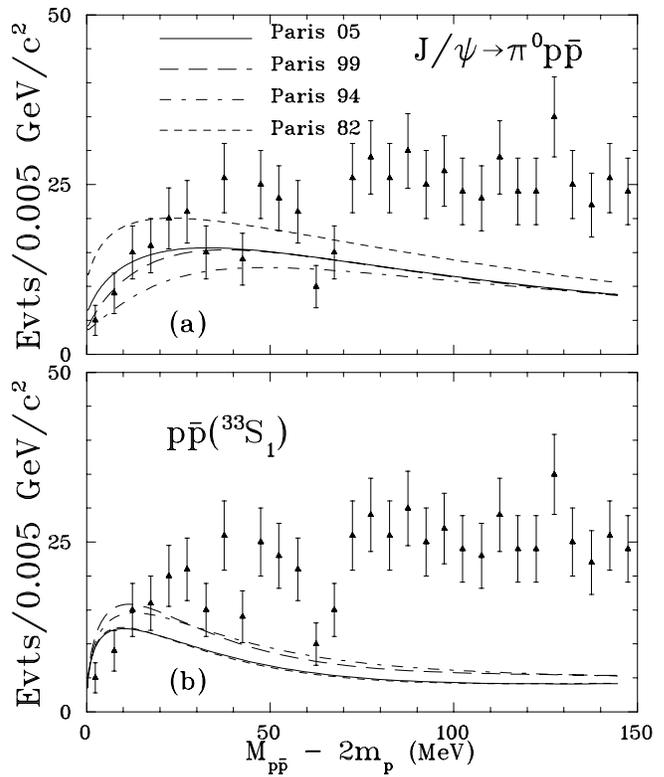}
\caption{ $ \pi^0  p \bar{p}  (^{33}S_1)$ decays. Data as in Fig.~1. 
(a) Final state factor $q \mid T_{ff}\mid^2  $
of Eq.~(\ref{watson}). 
(b) Rate  $q\mid T_{if}\mid^2  $  of Eq.~(\ref{pilkuhn}). 
This wave is not consistent with the  BES data 
whatever the    choice    of $p \bar{p}$ potential version, 
of $C_{if}$ in (a) or 
  $A_{if}(0)$ and   $r_o$ in (b).} 
\end{figure}

\begin{figure}
\label{fig3}
 \includegraphics*[width=8.5cm]{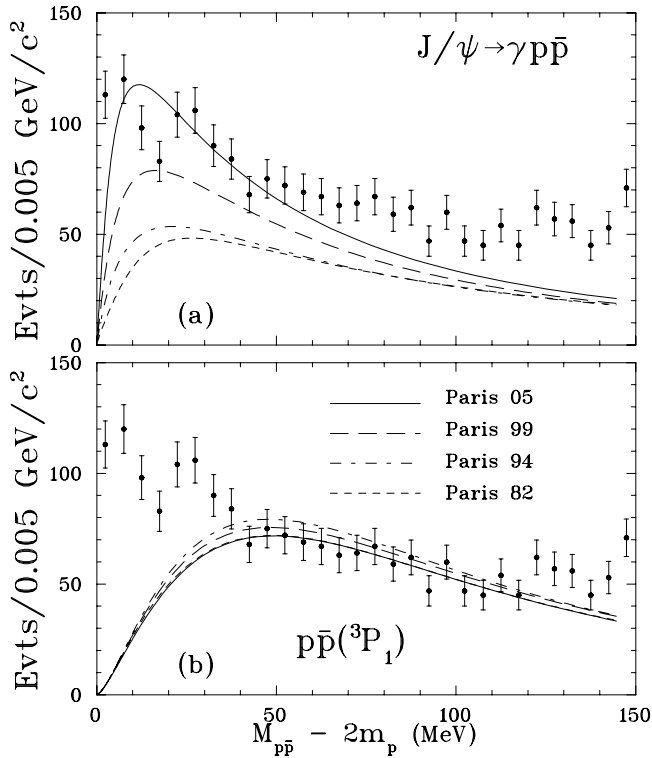}
\caption{$ \gamma  p \bar{p}  (^{3}P_1) $ decays.
Experimental data have been extracted from Fig.~3(a) 
of Ref.~\protect\cite{bai03}.(a) Final state factor 
$q \mid T_{ff}/q \mid^2$ of Eq.~(\ref{watson}). 
The enhancement in Paris 05 solution~\protect\cite{par05} is related 
to a 18~MeV wide state bound by 5~MeV in the $^{33}P_1$ wave.    
(b) Rate  $q\mid T_{if}\mid^2$  of Eq.~(\ref{pilkuhn}).
This wave cannot reproduce the  BES data 
whatever the  
 choice    of $p \bar{p}$ potential version, of $C_{if}$ in (a) or 
  $A^1_{if}(0)$ and   $r_o$ in (b).} 
\end{figure}

\begin{figure}
\label{fig4}
 \includegraphics*[width=8.5cm]{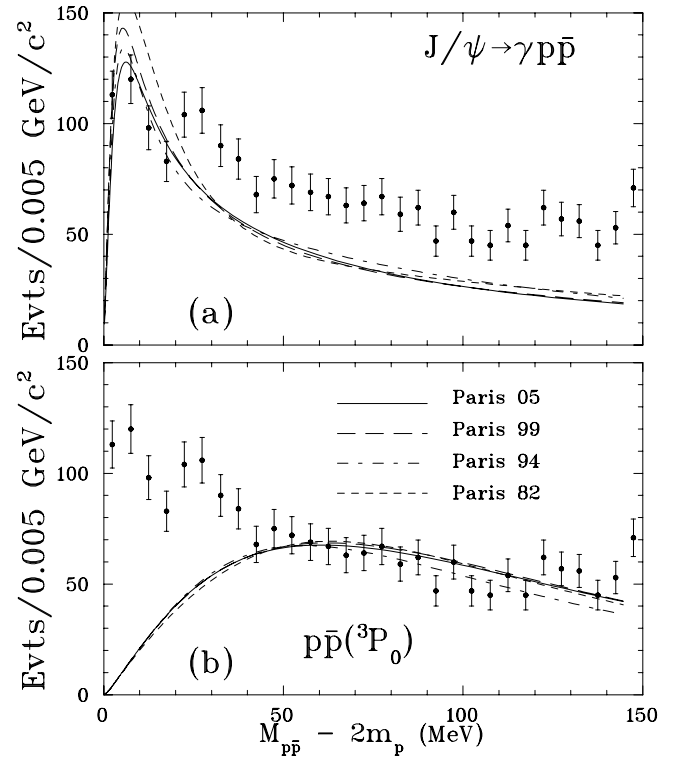}
\caption{ $ \gamma  p \bar{p}  (^{3}P_0) $ decays. Data as in Fig. 3.
(a) Final state factor $q\mid T_{ff}/q \mid^2$ of Eq.~(\ref{watson}).
The low energy part is   dominated  
by  the resonance  in the $ ^{13}P_0  $  
wave  at   $1876$~MeV, of  $10$~MeV width,  present in all models.    
However, for  $ q>1$~fm$^{-1}$ this approximation fails to fit   
the data.
(b) Rate  $q\mid T_{if}\mid^2$  of Eq.~(\ref{pilkuhn}) with  $r_o=0.55$~fm. 
This rate can describe only the    $q >1 $~fm$^{-1}$ part of the spectrum.
This wave is not consistent with the  BES data. } 
\end{figure}

\begin{figure}
\label{fig5}
 \includegraphics*[width=8.5cm]{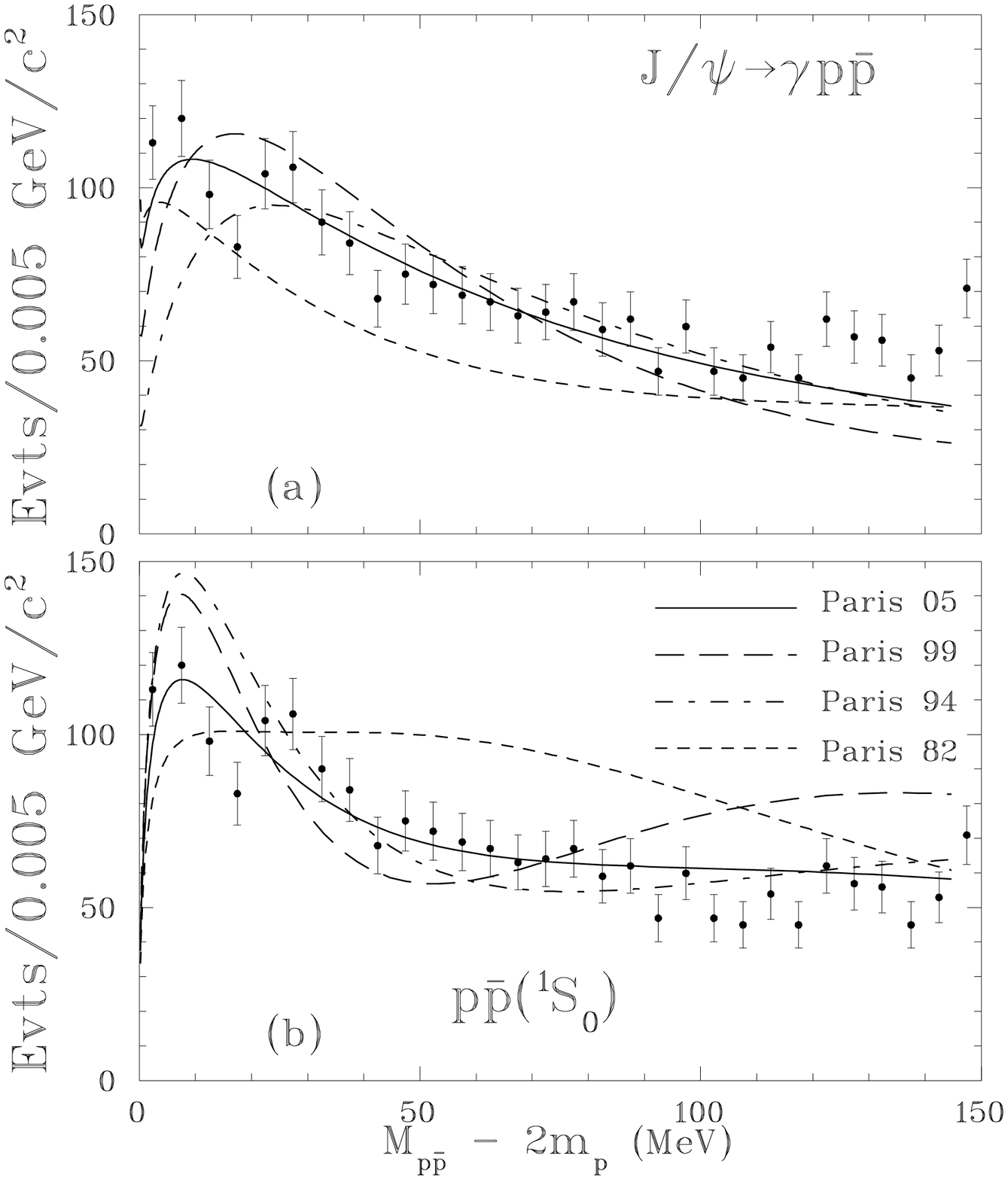}
\caption{ $ \gamma  p \bar{p}  (^{1}S_0) $ decays.  Data as in Fig. 3. 
(a)  Final state factor 
$q\mid T_{ff}\mid^2$ of Eq.~(\ref{watson}).    
At higher ($q >2$~fm$^{-1}) $ momenta  this approximation begins to fail.
(b) Rate\protect\linebreak $q\mid T_{if}\mid^2$ of Eq.~(\ref{pilkuhn}) with  $r_o=0.55$~fm. 
The latest  Paris  model~\protect \cite{par05} 
offers the best fit to the  data with an  $ ^{11}S_0 $ wave  
involving a  quasi-bound state  
located very close to threshold, of  $ 53 $~MeV width  and 
$ 5 $~MeV binding. } 
\end{figure} 

The advantage of K matrix formalism is clear in the analyzes 
of low-energy final state interactions  since it isolates  the kinematic 
singularity into a definite  form given by Eq.~(\ref{pilkuhn}). 
The two functions $ A_{ff}$ and $ A_{if} $  depend only on $ q^2$.
Hence,  close to the threshold, a  constant scattering length 
approximation  in  Eq.~(\ref{watson}) may well indicate some sub-threshold 
phenomena. This approximation has been used in Refs.~\cite{ker04,bug04}. 
In the $ p {\bar p}$ system the energy dependence in $ A_{ff}$  
is strong  as  pointed out in  Ref.~\cite{sib04} on the basis of an
one-boson exchange version of Bonn potential. A similar behavior 
is seen with the Paris 
model although these two potentials  differ  strongly in the two-pion sector.
 As shown below,   Eq.~(\ref{watson}) with a constant $ C_{if} $ and  realistic  
$ A_{ff}(q^2)$  describes a  too narrow energy range. 
The  selection of the best  $ p {\bar p}$  partial wave 
requires  Eq.~(\ref{pilkuhn}). This equation offers also an explicit and  unique 
dependence on the  the on-shell $ A_{ff}$ (or $ T_{ff}$ since 
$ 1/(1  + i q  A_{ff} ) = 1  -   i  q  T_{ff}$). 
An  off-shell $ A_{ff}$
may be involved in  $ A_{if} $ if one attempts to construct a model 
for the  $p {\bar p}$  formation.

\textit{Results}. 
There exists  substantial phenomenological control over  $ A_{ff} $.  
Here  these scattering lengths are calculated 
in terms of the   Paris $ N\bar{N}$  potential 
model and the same  procedure is  applied to both  decay modes 
 $ J/\psi \rightarrow  \pi ^0  p \bar{p}$ and 
$ J/\psi \rightarrow \gamma   p \bar{p} $. 
Figures 1 to  5  present the results obtained with Eq.~(\ref{watson}) 
and Eq.~(\ref{pilkuhn}) 
for the five   states of interest  
calculated for the  four  versions  of Paris  
model~\cite{par05,par99,par94,par82}  which evolved over the  last 20 years.
This evolution followed the increasing data basis. The last 
version~\cite{par05} is based on 3934 data which includes the recent  
antiproton-hydrogen widths and 
shifts~\cite{aug99,augnp99} and the total 
$ \bar{n}p$ cross sections of Ref.~\cite{iaz00}.
The Coulomb interactions  yield   
enhancements of the $S$ waves at very low energies due to Gamov  
factors. These affect  the 
final state interaction  
for $q < 0.15 $~fm$^{-1} $  and produce spikes.  Since the amplitudes are 
 weighted by the phase space factor $q$, these become  unessential. 
The $q$ factor represents a residual piece of the full three-body phase 
space~\cite{PDG02}. 
   

\textit{A qualitative model}.  
As exemplified  by the final state calculations,  the BES 
findings are most  consistent with a $  p \bar{p}(^{1}P_1) $ wave in the    
  $\pi^0   p \bar{p} $ channel and a   $  p \bar{p}(^{1}S_0) $ 
 wave in the  $\gamma   p \bar{p}  $ channel.     
Therefore the  experiment  leads us to a  simple 
picture for the  slow   $p \bar{p}$ formation. 

The  initial heavy  $c\bar{c}$ quarks in 
the $J/\psi$  state of   $  J^{PC} = 1^{--} $ 
annihilate into  a $N \bar{N} $ pair. 
As argued in Refs.~\cite{kro98,bro81} that process 
is  mediated by 
three gluon exchange.  Due to isospin conservation, the 
baryon pair is formed in an $I=0$ state of $n \bar{n} + p \bar{p} $
as indicated by experiment~\cite{ant93} and  calculations  
of Ref.~\cite{kro98}.  
The pair inherits the $J/\psi$ quantum numbers  
$  J^{PC} = 1^{--}$ and forms a  $^3S_1$ state.  
Next,  the emission of a  pion or a photon takes place. 
The $\pi_0$  emission  proceeds via the   
standard $ \pi N \bar{N}$ coupling 
$ (f_{\pi NN}/2m_{\pi})~\bf{q} \cdot \mbox{\boldmath$\sigma$}$.
It  requires one  nucleon to  flip spin   and  change angular momentum, which 
leads to  the final  $ p \bar{p} (^{1}P_1) $ state. 
The photon  may be produced as a magnetic one or as 
an electric one. The relevant formation amplitudes are given by 
the transition operator 
$  (e/2m_p)~ 
[ 2~\mbox{\boldmath$\epsilon$}^{\gamma} \cdot \bf{q} 
+ i~\mbox{\boldmath$\sigma$} \cdot (\bf{k}^{\gamma} \times  \mbox{\boldmath$\epsilon$}^{\gamma})]$. 
In the final states $q$  is small. In the intermediate 
states it is not necessarily so, but any formation mechanism 
would favor  small momenta.  Since $\mid \bf{k}^{\gamma}\mid $  is large 
we conclude that it is  the magnetic transition which is 
more likely to occur. 
It  favors  formation of the 
final $ \gamma  p \bar{p}(^1S_0) $ state which arises in 
a most natural way. In the initial $  p \bar{p}(^3S_1) $ wave, the proton 
and antiproton magnetic moments are
opposite and  the transition to  $ p \bar{p}(^1S_0)$ involves spin and 
magnetic moment flips. Large  moments  create  large radiative amplitudes.  
The  emission model indicated  above, yields 
comparable branching ratios of the $\gamma$ and $\pi^0$ channels, 
as found in the experiment.  This ratio follows roughly  
the ratio of the coupling constants 
$ f_{\pi NN}^2  / \left(4\ e^2\right)  \approx 2.8  $  
while the experimental ratio is  $\approx  3 $~\cite{PDG02}.

The final $  p \bar{p} $
state involves the isospin 1 plus isospin 0 combination. The pair may be also 
formed in the  $  n \bar{n} $ state  and undergo a transition 
to $  p \bar{p} $ in the final state. 
That process is expected to be suppressed, since  that  transition implicates  
the $ T_{ff}(I=1)- T_{ff}(I=0)$ amplitude 
which is about an order of magnitude smaller than the elastic 
 $ T_{ff}(I=1)+ T_{ff}(I=0)$ one.  
The simple model of final photon 
radiation discussed above would reduce   the neutron 
channel even further, due to different charges and magnetic moments.

\textit{Conclusions}. 
 We have shown that the new results  of the BES Collaboration 
find a natural explanation in a fairly traditional model 
of $ p \bar{p} $ interactions based on $G$-parity  transformation, 
dispersion theoretical  treatment of two pion exchange and  
semi-phenomenological  absorptive and short 
range potentials. This model predicts quasi-bound states close 
to the threshold,  in particular in the  
$ p \bar{p} (^{33}P_1) $  and  $ p \bar{p} (^{11}S_0)$ waves 
and a resonance in the $ p \bar{p} (^{13}P_0) $  wave. The first two 
 indicate a  strong dependence on the model parameters
and, so far,  are not  confirmed 
in other experiments.
The third one, the  resonant state, is well established.

It is the $^{1}S_0 $ state which   reproduces the 
$\gamma  p \bar{p} $  spectrum found by the  BES collaboration.  
This wave is dominated by a  strong attraction due to the  pion exchange 
forces. This attraction  generates  broad, deeply bound states.
The recent atomic and scattering data indicate that such a state 
 in the  $^{11}S_0$ wave is located close to the threshold. 
The BES data offer some support for the existence 
of such a state. 
The actual energy level and its width are affected 
by interactions at  distances less then 1~fm. These are not 
 not fully   understood and only partly controlled through phenomenology.  

In order to better see the nature of the  $^{11}S_0 $ 
state,  one should look directly  under the $p \bar{p} $ threshold.
This  could  be done with  measurements of the invariant mass 
of few meson systems  coupled to  $p \bar{p} $ just below the threshold.
The selectivity in partial waves is necessary, and a  convenient way 
to  reach  that is the  
$ J/\psi \rightarrow  \gamma$ $  mesons $ decay. 
Another,   indirect  method is to achieve a fine    
resolution of   energy levels in  antiprotonic atoms. 
Some  anomalies were  found in  atoms with  nuclei 
characterized by weakly bound  valence protons~\cite{trz01}.  
These  anomalies  may  reflect a   resonant behavior of  the 
$p \bar{p} $ scattering amplitudes  in the region of $p \bar{p} $ 
quasi-bound states.  More systematic measurements are necessary to 
pinpoint  the  $p \bar{p} $ wave 
responsible for these effects.

We acknowledge useful discussion on the angular distribution 
with J.-M.~Levy and W.~Kloet. We thank M.~Lacombe and B.~El-Bennich 
for helpful collaboration on the $p \bar{p} $ Paris potential. 
B.~L. is grateful to S.~Jin and 
P.~N.~Shen for important informations and enlightening comments. 
This work was performed in the framework of 
the IN2P3-Polish Laboratories Convention.


\begin{references}
\bibitem{kle02} E.~Klempt, F.~Bradamante, A.~Martin and J.-M.~Richard, 
Phys.~Rep. {\bf 368}, 119 (2002): 
\textit{Antinucleon-nucleon interaction at low energy: scattering and protonium}.

\bibitem{aug99}  M. Augsburger et al, Phys. Lett. \textbf{B461}, 417 (1999): 
\textit{Measurement of the strong interaction parameters in antiprotonic deuterium}. 

\bibitem{augnp99} M.~Augsburger et al.,  Nucl. Phys. {\bf A658}, 149 (1999): 
\textit{Measurements of the strong interaction parameters in antiprotonic hydrogen and probable evidence for an interface with inner bremsstrahlung}.

\bibitem{bai03} BES collaboration J.~Z.~Bai et al., 
 Phys. Rev. Lett. {\bf 91}, 022001 (2003): 
\textit{Observation of a Near-Threshold Enhancement in the $p\bar p$ Mass Spectrum from Radiative $J/\psi\to\gamma p\bar p$ Decays}.

\bibitem{got99} D.~Gotta et al., Nucl.  Phys. {\bf A660}, 283 (1999): 
\textit{Balmer $\alpha$ transitions in antiprotonic hydrogen and deuterium}.

\bibitem{par05} M.~Lacombe, B.~Loiseau, R.~Vinh~Mau, S.~Wycech,\\
\textit{The Paris $N\bar N$ potential constrained by recent $\bar np$ total cross section and antiprotonic-atom data}, in preparation.

\bibitem{par99} B.~El-Bennich, M.~Lacombe, B.~Loiseau, R.~Vinh~Mau,
Phys. Rev. C\textbf{59}, 2313 (1999): 
\textit{Refining the inner core of the Paris $N\bar N$ potential}.

\bibitem{par94} M.~Pignone, M.~Lacombe, B.~Loiseau, R.~Vinh~Mau,
Phys. Rev. C\textbf{50}, 2710 (1994): 
\textit{Paris $N\bar N$ potential and recent proton-antiproton low energy data}.


\bibitem{par82} J.~C\^{o}t\'e, M.~Lacombe, B.~Loiseau, B.~Moussallam, R.~Vinh~Mau, Phys. Rev. Lett. \textbf{48}, 1319 (1982): 
\textit{On the Nucleon-Antinucleon Optical Potential}.

\bibitem{sib04} A.~Sibirtsev, J.~Haidenbauer, S.~Krewald, Ulf-G.~Mei\ss ner 
and A.~W.~Thomas,Phys. Rev. D\textbf{71}, 054010 (2005): \textit{Near threshold enhancement of the $p \bar p$ mass spectrum in $J/\Psi$decay}.
 
\bibitem{ker04} B.~Kerbikov, A.~Stavinsky and V.~Fedotov, Phys. Rev. C{\bf 69}, 055205 (2004): 
\textit{Low-mass proton-antiproton enhancement: BELLE and BES results, premises of LEAR and expectations from CLAS}.

\bibitem{bug04} D.~V.~Bugg, Phys. Lett. {\bf B598}, 8~(2004): 
\textit{Reinterpreting several narrow 'resonances' as threshold cusps}.

\bibitem{gao04} Chong-Shou~Gao and Shi-Lin~Zhu, Commun.~Theor. Phys. {\bf 42}, 844~(2004): 
\textit{Understanding the Possible Proton Antiproton Bound State Observed by BES Collaboration}.

\bibitem{zou04} B.~S.~Zou, H.C.~Chiang, 
 Phys. Rev. {\bf D69}, 034004 (2004): 
\textit{One-pion-exchange final-state interaction and $p\bar p$ near threshold enhancement in $J/\psi\to\gamma p\bar p$ decays}.

\bibitem{PDG02} Particle Data Group,  S.~Eidelman et. al.,
\textit{Review of Particle Physics}, Phys. Lett. {\bf B592}, 1~(2004). 

\bibitem{gur71} M.~Gourdin, in \textit{Hadronic interactions of electrons 
and photons}, p.~395, Acad. Press 1971, Ed. J.~Cumming and H.~Osborn. 

\bibitem{ren86} F.~M.~Renard, \textit{Basics of electron positron collisions}, 
\'Editions Fronti\`eres, 1981, p.~116.

\bibitem{pil03} H.~Pilkuhn, \textit{ Interaction of Hadrons}, North Holland 
P.~C., 1967, p.~213, Eq.~3.7.

\bibitem{kro98} J.~Bolz and P.~Kroll, Eur.~Phys.~J. \textbf{C~2}, 545 (1998): 
\textit{Exclusive $J/\Psi$ and $\Psi'$ decays into baryon-antibaryon pairs}.

\bibitem{iaz00} OBELIX Collaboration, F.~Iazzi et al.,
\textit{Phys. Lett.} \textbf{B475}, 378 (2000): 
\textit{Antineutron-proton total cross-section from 50 to 400 MeV/c}.

\bibitem{bro81} S.~J.~Brodsky and G.~P.~Lepage, Phys. Rev. D{\bf 24}, 2848 (1981): 
\textit{Helicity selection rules and tests of gluon spin in exclusive quantum-chromodynamic processes}.

\bibitem{ant93} FENICE Collaboration,  A.~Antonelli et al., 
              Phys. Lett.  {\bf B301}, 317 (1993): 
\textit{A new measurement ot $J/\Psi \to n\bar n$}. 

\bibitem{trz01}  A.~Trzcinska et al. Nucl. Phys. {\bf A692}, 176c (2001): 
\textit{Information on anti-protonic atoms and the nuclear periphery
from the PS209 experiment};
 S.~Wycech, Nucl. Phys. {\bf A692}, 29c (2001): 
\textit{Optical potential for anti-proton nucleus interactions}.

\end{references}
\end{document}